\documentclass[pre,twocolumn,groupedaddress,nopacs]{revtex4}

\usepackage{graphicx}

\begin{document}

\title{Fast Simulation of Facilitated Spin Models}

\author{Douglas J. Ashton, Lester O. Hedges and Juan P. Garrahan}

\address{School of Physics and Astronomy, University of Nottingham,
Nottingham, NG7 2RD, UK}

\begin{abstract}
We show how to apply the absorbing Markov chain Monte Carlo algorithm
of Novotny to simulate kinetically constrained models of glasses.  We
consider in detail one-spin facilitated models, such as the East
model and its generalizations to arbitrary dimensions.  We investigate
how to maximise the efficiency of the algorithms, and show that
simulation times can be improved on standard continuous time Monte
Carlo by several orders of magnitude.  We illustrate the method with
equilibrium and aging results.  These include a study of relaxation
times in the East model for dimensions $d=1$ to $d=13$, which provides
further evidence that the hierarchical relaxation in this model is
present in all dimensions.  We discuss how the method can be applied
to other kinetically constrained models.
\end{abstract}


\maketitle

\section{Introduction}

By nature, glassy systems are dynamically slow, with relaxation times
found to increase rapidly as the temperature $T$ is lowered
\cite{Angell95,DebenedettiS01,Ediger96}.  One consequence of this slow
down is the difficulty in probing the dynamics of such systems in the
long time regime through means of numerical simulation.  Traditional
numerical methods often become insufficient to obtain data within a
realistic time frame thanks to time scales growing at best
exponentially with inverse temperature and in most cases much faster.
A common algorithm used for systems close to dynamical arrest is the
rejection free algorithm known as the ``n-fold way'', or continuous
time (CT) Monte Carlo \cite{BortzKL75,NewmanBarkema}.  While CT can
provide an impressive improvement over standard Monte Carlo (MC), it
can still become inefficient for some extremely slow systems.

The natural generalization of the CT algorithm is Novotny's Monte
Carlo with Absorbing Markov Chains (MCAMC) \cite{Novotny95}, which so
far has been mainly used to study magnetic reversal.  In this paper we
show how to apply the MCAMC method to simulate kinetically constrained
models (KCM) of glass formers \cite{RitortS03}, in particular,
facilitated spin models such as the Fredrickson-Andersen (FA) model
\cite{FredricksonA84} and the East model \cite{EisingerJ93} in
arbitrary dimensions.  We show that in the case of East models the
MCAMC method allows to improve simulation times, both for equilibrium
and out-of-equilibrium dynamics, by several orders of magnitude at low
temperatures.

The paper is organized as follows.  In Section \ref{mcamc} we outline
the MCAMC technique.  Section \ref{east} provides a detailed analysis
of its application to the one-dimensional East model.  In Section
\ref{approx} we discuss several important approximations which allow
to maximise the computational gain.  The method is generalized to the
East model in arbitrary dimensions in Section \ref{alld}, and, in
Section \ref{xover}, to models that interpolate between the East and
FA cases \cite{BuhotG01}.  Section \ref{highers} describes faster,
higher level versions of MCAMC.  In Section \ref{tests} we present
speed tests of the algorithms.  In Section \ref{results} we show some
illustrative results obtained with the MCAMC method, including a study
of the $T$ dependence of relaxation times in the East model for
dimensions $d=1$ to $d=13$.  We conclude in Section \ref{conclu}
discussing the possible implementation of the method to simulate other
KCMs.

\section{Outline of the MCAMC Method}
\label{mcamc}

Continuous time Monte Carlo works well for systems where the vast
majority of moves are most likely to be rejected due to kinetic or
energy constraints \cite{NewmanBarkema}.  In order to avoid constantly
attempting and failing to make a move, as in standard MC, in CT one
fast-forwards the system, always accepting moves, and updating the
clock by how long it would have taken in standard MC.  This can speed
things up greatly in systems with slow dynamics because making an
unfavourable move can take a long time.

The trouble with most slow systems is that often after an unlikely
move, even in the CT scheme, the most likely move is to undo what has
just been done before.  With the MCAMC method of Novotny
\cite{Novotny95}, instead of just fast forwarding to when one is
accepted, one can fast-forward to when two (or more) unlikely moves
have been accepted, updating the clock appropriately.  To do this
whilst keeping the dynamics exact and keeping to detailed balance one
must use the formalism of absorbing Markov chains \cite{Novotny95}
(see \cite{Novotny01} for a pedagogical review).

In an MC algorithm any given move depends only on the two states that
the system is moving between and not on any previous moves.  This is
by definition a Markov process and allows us to treat the MC algorithm
as a Markov chain \cite{GrimmettStirzaker}.  A Markov chain is
characterised by the matrix $\mathbf{M}$ which defines transition
probabilities between states.  If the vector $\vec{x}^T(m)$ indicates
the probability distribution of the system after iteration $m$, the
probability distribution at the next step $m+1$ is given by
$\vec{x}^T(m+1)=\vec{x}^T(m)\mathbf{M}$.

We define an absorbing Markov chain by separating the available states
into $s$ transient states and $r$ absorbing ones
\cite{Novotny95,Novotny01}. The system always starts in a transient
state and by successive applications of the Markov matrix explores the
transient subspace until it lands in an absorbing (or exit) state from
where it cannot leave.  We can divide the general state vector
$\vec{x}^T$ into absorbing and transient parts, to get the
$(r+s)$-dimensional vector $\vec{x}^T = ( \vec{u}^T , \vec{v}^T )$
where $\vec{v}^T$ contains the transient states.  The initial state in
this form must obey $\vec{x}_I^T = ( \vec{0}^T , \vec{v}_I^T )$.  With
this structure the Markov matrix can be written in the form,
\begin{equation}\label{generalM}
\mathbf{M} = \left( \begin{array}{cc} \mathbf{I}_{r \times r} &
\mathbf{0}_{r \times s} \\ \mathbf{R}_{s \times r} & \mathbf{T}_{s
\times s} ,
 \end{array} \right)
\end{equation}
where $\mathbf{I}$ is the identity matrix, $\mathbf{0}$ is the
zero matrix and subscripts indicate the size of each matrix.  The
positions of the identity and zero matrices guarantee that if the
system falls into an absorbing state then it does not leave.  The
transient matrix, $\mathbf{T}$, gives the probabilities for moving
between transient states and the recursive matrix, $\mathbf{R}$,
gives the probabilities for moving from the transient states to
the absorbing states.

For a given starting vector $\vec{v}_I^T$ the probability of still
being in the transient subspace, $p_{\rm trans.}$, after $m$ steps is
\begin{equation}\label{prob still in transient space}
p_{\rm trans.}=\vec{v}_I^T \mathbf{T}^m \vec{e} ,
\end{equation}
where $\vec{e}$ is a vector with all elements equal to 1.  The
probability of absorbing to a particular state after $m$ steps is
given by summing over the probabilities of absorbing at each time
step.  This gives the  vector $ \vec{p}^T_{\rm abs.}$,
\begin{equation}\label{absorption after m steps}
    \vec{p}^T_{{\rm abs. after}~m}=\vec{v}_I^T\bigl( \mathbf{I} +
    \mathbf{T}+ \cdots + \mathbf{T}^{m-1}\bigr)\mathbf{R} .
\end{equation}
If the exit has taken place at step $m$, then the probabilities of
absorbing into the different exit states is given by:
\begin{equation}
\label{pm}
\vec{p}^T_{{\rm abs. at}~m} = \frac{\vec{v}_I^T \mathbf{T}^{m-1}
  \mathbf{R}}{\vec{v}_I^T \mathbf{T}^{m-1} \mathbf{R} \vec{e}} .
\end{equation}

Here it is convenient to introduce the fundamental matrix
\begin{equation}\label{fundamental matrix def}
    \mathbf{N} = (\mathbf{I}-\mathbf{T})^{-1} = \mathbf{I} +
    \mathbf{T} + \mathbf{T}^2 + \cdots ,
\end{equation}
which can be used to obtain the probability that the system will
absorb to a particular state irrespective of when it exits,
\begin{equation}\label{prob absorb particular state}
    \vec{p}^T_{\mathrm{abs.}}=\vec{v}_I^T\mathbf{NR} .
\end{equation}
The fundamental matrix can also be used to determine the average time
to leave the transient subspace \cite{Novotny95,Novotny01}
\begin{equation}\label{average exit time}
    \langle\tau\rangle=\vec{v}_I^T \mathbf{N} \vec{e} .
\end{equation}

Once our system is in the initial state we can generate an exit time
by solving the inequality
\begin{equation}\label{update time inequality}
    \vec{v}_I^T \mathbf{T}^m \vec{e} < \overline{r} \leq \vec{v}_I^T
    \mathbf{T}^{m-1} \vec{e} ,
\end{equation}
where $\overline{r}$ is a random number between 0 and 1.  Next, we use
a second random number to choose an absorption state from the
distribution in equation (\ref{pm}) and then we can update the system
appropriately.  The new state will become the initial state in another
absorbing Markov chain, and so on.  A successful MCAMC algorithm will
choose transient states such that the system tends to move between
them many times before exiting.

\section{Application of MCAMC to Kinetically Constrained Models}
\label{east}

The MCAMC algorithm has been used in the study of magnetic reversal
and nucleation in systems such as the Ising model, see
e.g. \cite{Novotny95,ParkRBN04,BuendiaRPN04}.  In the case of magnetic
reversal, for example, there is a well defined initial transient state
corresponding to the metastable configuration in which all spins are
aligned in one direction.  This state represents a deep minimum in the
energy: on reversing a single isolated spin the overwhelming
likelihood is that the system will immediately relax back to the
initial state.  An MCAMC algorithm with two transient states ($s=2$)
overcomes this problem: the system can escape from the metastable
configuration by insisting that two spins are flipped simultaneously;
this process is characterised by two transient states, the metastable
configuration and that corresponding to a single spin flip
\cite{Novotny95}.

Kinetically constrained models (KCMs) are frequently used to study
the dynamical behaviour of glass formers \cite{RitortS03}.  Of
particular interest is the analysis of relaxation times and the
emergence of length-scales \cite{GarrahanC02}.  Similarly to the
magnetization reversal problem above, at low temperatures or high
densities, these systems evolve by falling into deep energy or
free energy traps from which it is difficult to escape.  The
difference, however, is that there is a multiplicity of trapping
states, not just one as in the Ising model case, which change as
the system evolves dynamically. In order to apply the MCAMC method
to KCMs one needs to identify the nature of the transient states
in these systems.

As an example consider the East model in one-dimension
\cite{EisingerJ93,RitortS03}.  The East model consists of a chain of
Ising spins, $n_i=\{0;1\}$, upon which a directional facilitation
constraint is imposed: a given spin $n_i$ may only flip if its nearest
neighbour to the left is in the excited state, $n_{i-1}=1$.  The rates
for allowed moves are such that the equilibrium distribution is that
corresponding to the energy function $H=J \sum_i n_i$ (we set $J=1$
from now on),
\begin{equation}
10 \stackrel{\epsilon}{\longrightarrow} 11, \,\,\,
11\stackrel{1}{\longrightarrow} 10, \,\,\ \label{East model rates}
\end{equation}
\noindent where $\epsilon \equiv e^{-\beta}$ and $\beta \equiv 1/T$.
 The result is a model in which excitations propagate in the eastward
 direction. The East model has been found to reproduce the dynamic
 behaviour of fragile glassy materials
 \cite{RitortS03,GarrahanC03,BerthierG05}.

From the rate equations above one can see that it is energetically
 unfavourable to excite a down spin the process becoming increasingly
 unlikely as temperature is decreased.  Consider a configuration with
 only isolated excitations:
\[\cdots100\cdots100\cdots100\cdots\]
When in this state the only choice is to excite a spin and pay the
accompanying energy penalty.  Consequently it is highly likely that
the raised spin is immediately relaxed, hence returning to the
previous state.  This is analogous to the all-up configuration of the
Ising model example.  It is important to note that in this isolated
state all excitations must be separated by a minimum of two unexcited
spins.  When separated by only a single spin there are two possible
outcomes from the creation of an excitation, i.e. one can create a
double state $110$ or a triplet $111$.  This produces an unnecessary
complication since the algorithm can no longer be classified by two
simple transient states.

By analogy to the Ising model one may define two transient states for
the system, the isolated configuration described above and states in
which a single excitation pair exists.  However, unlike the case of
magnetic reversal, it is clear that neither of the transient states
identified for the East model are unique. It is possible to construct
numerous configurations which satisfy the above criterion, in essence
we have identified two classes of transient state. Once again the
absorbing states consist of all configurations attainable by the
excitation of two spins, either forming two isolated doubles or a
triplet state,
\[
\cdots110\cdots110\cdots100\cdots
\]
\[
\cdots111\cdots100\cdots100\cdots
\]
For the East model it is possible to classify each lattice site
according to its local neigbourhood. Taking a site along with its
nearest and next-nearest neighbour to the right, each site can be
classed according to a binary labelling scheme, i.e. $100 \equiv 4$,
$110 \equiv 5$, etc., where the number of sites in each class is,
$N_4$, $N_5$, etc.  Using this notation we define the entry condition
for the algorithm with $s=2$ transient states as the point at which
the number of sites in class 4 equals the total number of excitations
present in the lattice, $M$, i.e. $N_4 = M$.

Before constructing the transient and recursive matrices it is
necessary to determine the probabilities for all possible transitions
between the different states. The transient and recursive states may
be labelled as follows,
\[\cdots100\cdots100\cdots\hspace{1 cm} v_1\]
\[\cdots110\cdots100\cdots\hspace{1 cm} v_2\]
\[\cdots110\cdots110\cdots\hspace{1 cm} u_1\]
\[\cdots111\cdots100\cdots\hspace{1 cm} u_2\]
with the following transition probabilities
\begin{eqnarray*}
P(v_1 \rightarrow v_2) &=& \frac{\epsilon N_4}{N} , \\
P(v_2 \rightarrow v_1) &=& \frac{1}{N} , \\
P(v_1 \rightarrow u_1) &=& 0 , \\
P(v_1 \rightarrow u_2) &=& 0 , \\
P(v_2 \rightarrow u_1) &=& \frac{\epsilon (N_4-1)}{N} , \\
P(v_2 \rightarrow u_2) &=& \frac{\epsilon}{N} ,
\end{eqnarray*}
where $N$ is the system size.

These transition probabilities are then used to build the transient
and recursive matrices for the system
\begin{eqnarray}
  \mathbf{T} &=& \left(
\begin{array}{cc}
  1-x & x \\
  y & 1-x-y\\
\end{array}
\right)
\label{Ts2}
, \\ \mathbf{R} &=& \left(
\begin{array}{cc}
  0 & 0 \\
  x-\epsilon y & \epsilon y\\
\end{array}
\right) ,
\end{eqnarray}
where $x=\frac{\epsilon N_4}{N}$ and $y=\frac{1}{N}$.

The absorption probabilities for the $u_1$ and $u_2$ states can be
found by taking the fundamental matrix, $\mathbf{N}$, and solving
equation (\ref{prob absorb particular state}) giving
\begin{eqnarray}
  P(u_1) &=& 1-\frac{1}{N_4}\label{prob double} , \\ P(u_2) &=&
  \frac{1}{N_4} ,
\label{prob triple}
\end{eqnarray}
where we have used an initial state vector $\vec{v}_I^T = (1\ 0)$.

To determine the exit time one must choose a random number and
iteratively solve the inequality given in equation (\ref{update
time inequality}). One then proceeds to choose an exit state from
the distribution formed by the exit probabilities, equations
(\ref{prob double}) and (\ref{prob triple}). It is clear that
matrices $\mathbf{T}$ and $\mathbf{R}$ are characterised by the
variable $N_4$ and as such both the probability distribution for
the absorption states and the exit time are governed by the entry
state, each state having its own unique solution.

This $s=2$ construction provides an algorithm that improves on
standard continuous time, $s=1$, by a factor proportional to
$e^\beta/N_4$.  This improvement in computational speed is offset
by the algorithmic complexity required to formulate the $s=2$
model.

\section{Approximations for the Update Time} \label{approx}

Computationally, the most expensive part of the algorithm as described
above is the procedure used to determine the time to exit from the
transient state. To perform the calculation exactly involves
diagonalising the $\mathbf{T}$ matrix and iteratively solving the
inequality using the halving method \cite{NewmanBarkema} or something
similar.  There are, however, a number of approximations that we can
employ to get around this. The exact form for
$\vec{v}_I^T\mathbf{T}^m\vec{e}$ for the $s=2$ case is
\begin{equation}\label{exact vtme}
    \vec{v}_I^T \mathbf{T}^m \vec{e} = \frac{1}{2} \left[
    \lambda_2^m+\lambda_1^m
    -(\lambda_2^m - \lambda_1^m) \left( \frac{1}{4z}+z \right) \right] ,
\end{equation}
where $z=\frac{1}{2}\sqrt{1+4\epsilon N_4 }$ and $\lambda_1$,
$\lambda_2$ are the eigenvalues of $\mathbf{T}$.  Both eigenvalues are
quite close to (and less than) 1.  However, in the limit of large $m$,
we have that $(\lambda_2 / \lambda_1)^m\ll1$, allowing us to simplify
equation (\ref{exact vtme}).  If we drop the restriction that $m$ must
be discrete then we can write (\ref{update time inequality}) as an
equality,
\begin{equation}\label{approx m small lamda}
    m \approx  \log \left(
    \frac{2\overline{r}}{1+z+1/4z} \right) \bigg{/} \log (\lambda_1) ,
\end{equation}
where again $\overline{r}$ is a random number between 0 and 1.
Both $z$ and $\lambda_1$ depend on $N_4$ and can be stored in a
lookup table.  The approximation works best when $m$ is large, so
for low temperatures ($T < 1$) where the time steps are larger it
works very well.  At higher temperatures one must be careful using
this approach.

Another possibility is to free oneself from the requirement to pick
the update time from a distribution and use instead the average. This
does mark a departure from the exact Monte Carlo algorithm, but in
most cases it turns out to be a reasonable simplification (it is
analogous to the approximation made when going from the n-fold
algorithm \cite{BortzKL75} to the CT one \cite{NewmanBarkema}).  If we
take the average time, then we can use equation (\ref{average exit
time}) which requires calculation of the fundamental matrix,
$\mathbf{N}$, either analytically or numerically.  For the East model,
$\mathbf{N}$ only depends on the number of excitations $M$, so the
time updates can be stored in a lookup table allowing for a
significant increase in speed.

To check the validity of using the average value for time updates
instead of picking them from a distribution, we can use the result
\begin{equation}\label{lifetime moment}
    \langle\tau^2\rangle=\vec{v}_I^T
    \bigl(2\mathbf{N}^2-\mathbf{N}\bigr) \vec{e} ,
\end{equation}
with (\ref{average exit time}) to calculate the mean square
fluctuations.  This shows that for lower temperatures the error on any
given measurement is $\sim\langle\tau\rangle$. Whilst this seems large
it is important to remember that we are always looking at logarithmic
time and on this axis the error is less significant. Also there are
many iterations between sampling points and the measurements are
averaged over many runs which will help to reduce any discrepancy. All
the simulations for this paper were performed using the average time
update.

\section{Generalisation to Any Dimension} \label{alld}

The method described in the previous section can easily be extended
allowing one to construct generalised transient and recursive matrices
for the East model in any spatial dimension $d$.  Considering the
transient states for the system it is clear that the $s=2$ algorithm
is triggered when all excitations within the lattice are isolated by a
region of space which encompasses all moves attainable by two
successive spin flips.  The $d=2$ analog of the ``100'' above is:
\begin{eqnarray*}
100\equiv\
\begin{array}{ccc}
  0 &  &  \\
  0 & 0 &  \\
  1 & 0 & 0 \\
\end{array}
\end{eqnarray*}
where no triangles may overlap if the algorithm is to trigger. As
for the $d=1$ case, the $\mathbf{T}$ and $\mathbf{R}$ matrices are
obtained by evaluating the probabilities for all transitions
between the transient and absorbing states. This analysis yields
matrices of the following form
\begin{eqnarray}
  \mathbf{T} &=& \left(%
\begin{array}{cc}
  1-x d & x d \\
  y & 1-y-x d - (d-1)\epsilon y\\
\end{array}%
\right) \\
  \mathbf{R} &=& \left(%
\begin{array}{cc}
  0 & 0 \\
  x d-\epsilon y & \epsilon y d\\
\end{array}%
\right)
\end{eqnarray}
It is easy to solve for the exit probabilities to each of the
absorbing states, where $N_4$ again indicates the number of
excitations in the isolated state,
\begin{eqnarray}
  P(u_1) &=& 1-\frac{d}{d-1+N_4 d}\\ P(u_2) &=& \frac{d}{d-1+N_4 d} .
\end{eqnarray}
The average lifetime to exit from the transient subspace may be
obtained from (\ref{average exit time}),
\begin{equation}
    \langle\tau\rangle = \frac{e^{2\beta} + (2N_4 d + d - 1)e^\beta}{N_4 d (N_4 d + d - 1)} .
\end{equation}

For systems in which $\beta$ is high and $N_4$ is large compared to
$d$, one may approximate the exit time to
\begin{equation}\label{average exit time general}
    \langle\tau\rangle \approx \frac{e^{2\beta}}{(N_4 d)^2} ,
\end{equation}
hence time steps are a factor of $d^2$ smaller than those of the East
model in $d=1$.

\section{FA-East Crossover model}
\label{xover}

The MCAMC algorithm described above can also be applied to the FA
model \cite{FredricksonA85,RitortS03} and, more generally, to the
model that interpolates between the FA and East models
\cite{BuhotG01}, which serves as a simple model for fragile-to-strong
transitions.  This model is characterised by the rates
\begin{equation}
11 \stackrel{b}{\longrightarrow} 01, \,\,\,
01\stackrel{b\epsilon}{\longrightarrow} 11, \,\,\,\
11\stackrel{(1-b)}{\longrightarrow} 10, \,\,\,\
10\stackrel{(1-b)\epsilon}{\longrightarrow} 11 . \,\,\
\label{FA-East model rates}
\end{equation}
The limit $b\rightarrow 0$ corresponds to the East model, and the
limit $b\rightarrow 1/2$, to the FA model.  For intermediate values of
$b$ the model displays a crossover between East-like dynamics at
higher temperature, to FA dynamics at low temperature.

The MCAMC algorithm is applied in much the same way as in the East
model case, except that now we have to allow for the possibility of
movement to the {\em west}.  In the simplest version, the transient
states are the same as for the East model, and the absorption states
are increased to include any move to the left. The result is an $s=2$,
$r=5$ absorbing Markov chain, with the following transient states
\[0100\cdots0100\cdots\hspace{1 cm} v_1, \]
\[0110\cdots0100\cdots\hspace{1 cm} v_2, \]
and absorbing states
\[0110\cdots0110\cdots\hspace{1 cm} u_1, \]
\[0111\cdots0100\cdots\hspace{1 cm} u_2, \]
\[1100\cdots0100\cdots\hspace{1 cm} u_3, \]
\[0110\cdots1100\cdots \mathrm{\ \ \  or \ \ \ } 1110\cdots0100\cdots\  u_4, \]
\[0100\cdots0100\cdots\hspace{1 cm} u_5. \]
The two states in $u_4$ are the same as far as this algorithm is
concerned.  Caution would be required if they were to be used as
transient states.  The transient and recursive matrices are as follows
\begin{equation}\label{FA-East transient matrix}
    \mathbf{T} = \left(%
\begin{array}{cc}
  1-xd & xda \\
  ya & 1-y-xd-a(d-1)\epsilon y \\
\end{array}%
\right)
\end{equation}

\begin{equation}\label{FA-East recursive matrix}
    \mathbf{R}=\left(%
\begin{array}{ccccc}
  0 & 0 & bxd & 0 & 0 \\
  a(xd-\epsilon y) & a\epsilon dy & 0 & bxd & by \\
\end{array}%
\right)
\end{equation}
where $a \equiv 1-b$, and $x$ and $y$ are the same as for the East
model case.

From here on the procedure is exactly the same as with the East
model except that there are more absorbing states to choose from.
One may use Eq. (\ref{prob absorb particular state}) to obtain
values for the absorption probabilities.  Caution is required as
the approximation breaks down in the regime of high temperature
and high symmetry ($b \rightarrow 1/2$). It should be noted that,
while less striking, even in the FA limit of $b=1/2$ this
algorithm outperforms standard CT.

\section{Higher order MCAMC}
\label{highers}

The entirely isolated state is problematic in terms of the dynamics of
the East model.  In order to relax the isolated excitations must
propagate in the lattice until they encounter another excitation along
the direction of facilitation.  Movement of this nature is promoted by
the occurrence of branching events,
\[100 \rightarrow \cdots \rightarrow 111 \rightarrow 101 .\]

The ``triplet'' absorption state, $u_2$, is the rate limiting step for
branching events, and hence the propagation of excitations in the
lattice.  However, from the absorption probabilities, Eqs.\ (\ref{prob
double}) and (\ref{prob triple}), we see that compared to the $u_1$
state the $u_2$ exit state is suppressed by a factor of
$\frac{1}{N_4-1}$.  Hence, the formation of triplets is unlikely,
particularly when the system size is large.

To overcome this problem it is possible to extend the MCAMC algorithm
to include the $u_1$ state as a transient state of the system.  There
are now three transient and three absorbing states, one absorbing
state corresponds to the $u_2$ state of the $s=2$ algorithm, the
remainder corresponding to configurations attainable from the new
transient state.  In one dimension one can represent the states as
follows,
\[\cdots100\cdots100\cdots100\cdots\hspace{1 cm} v_1, \]
\[\cdots110\cdots100\cdots100\cdots\hspace{1 cm} v_2, \]
\[\cdots110\cdots110\cdots100\cdots\hspace{1 cm} v_3, \]
\[\cdots110\cdots110\cdots110\cdots\hspace{1 cm} u_1, \]
\[\cdots111\cdots100\cdots100\cdots\hspace{1 cm} u_2, \]
\[\cdots111\cdots110\cdots100\cdots\hspace{1 cm} u_3. \]

Once again the transient and recursive matrices may be constructed
by considering all possible transitions between the states.
\begin{eqnarray}
  \mathbf{T} &=&  \left(%
\begin{array}{ccc}
  1-x & x & 0 \\
  y & 1-x-y & x-\epsilon y \\
  0 & 2y & 1-2y-x \\
\end{array}%
\right)\\
  \mathbf{R} &=& \left(%
\begin{array}{ccc}
  0 & 0 & 0 \\
  0 & \epsilon y & 0 \\
  x-2\epsilon y & 0 & 2\epsilon y \\
\end{array}%
\right) .
\end{eqnarray}
The $s=2$ transient matrix, Eq.\ (\ref{Ts2}), is now a submatrix of
$\mathbf{T}$; the addition of an extra transient state has appended
one extra row and column to the matrix, the rest of the structure
remaining intact.

Unlike the case of $s=2$, it is not so simple to generalise the $s=3$
matrices for any dimension. This arises from the non-equivalence of
the $v_3$ state in dimensions $d>1$, i.e. in two dimensions
\begin{eqnarray*}
\begin{array}{ccc}
  0 &  &  \\
  1 & 0 &  \\
  1 & 1 & 0 \\
\end{array}%
\neq
\begin{array}{ccc}
  0 &  &  \\
  1 & 0 &  \\
  1 & 0 & 0 \\
\end{array}%
\cdots
\begin{array}{ccc}
  0 &  &  \\
  1 & 0 &  \\
  1 & 0 & 0 \\
\end{array}%
\end{eqnarray*}
When considered as absorption states the two configurations above may
be treated identically since the probability of exiting to each state
is the same.  However, as transient states each configuration has
different exit probabilities and as such must be treated
independently.  In essence, one requires an $s=4$ algorithm to provide
the equivalent result in dimension two and above.

Returning to the $d=1$ example, we find that the $u_2$ is now the most
likely absorption state.  This is because all other exit states
require the excitation of an additional spin, i.e., they are
suppressed by a factor of $e^\beta$.  Solving for the average lifetime
gives
\[
\langle\tau\rangle \approx \frac{e^{2\beta}}{N_4}.
\]
Hence, $s=3$ improves on $s=2$ by a factor of $N_4$.  While this may
seem a modest enhancement in performance, note that the extra
algorithmic complexity required to develop $s=3$ is negligible.
Having made a working $s=2$ algorithm one may essentially use $s=3$
for free.

As CT enables one to obtain an $e^\beta$ speed increase over
traditional MC, $s=2$ enables one to achieve a further improvement of
$e^\beta$ over CT.  In effect, $s=2$ enables one to bypass all
$e^\beta$ processes (i.e. those that involve the excitation of a
single spin) by insisting that two successive spins are excited. The
double and triplet states of the $s=2$ model are examples of
$e^{2\beta}$ processes.  In order to construct an algorithm with a
further $e^\beta$ speed gain requires one to identify all of the
$e^{2\beta}$ arrangements and include them as transient states. This
means that the absorption states now correspond to all configurations
attainable from the transient states which result in the simultaneous
excitation of three spins. This analysis leads to an $s=7$ algorithm
consisting of seven transient and seven absorbing states.

In one dimension, $s=7$ may be triggered when all excitations within
the lattice are separated by at least three unexcited spins,
i.e. $1000\cdots1000$.  To maximise performance it is useful to use a
hybrid algorithm consisting of $s=1$, $3$ and $7$ components with each
sub-algorithm activated by its own triggering condition.

In order to improve algorithmic efficiency it is convenient to
compute absorption probabilities using Eq. (\ref{prob absorb
particular state}) rather than the exact form of Eq. (\ref{pm}).
Unlike the case of $s=2$, where it may be shown that two
expressions are identical, for higher order algorithms the
solution of Eq. (\ref{prob absorb particular state}) only provides
an approximation. In general one must employ caution when using
this approach. For both $s=3$ and $s=7$ it has been shown that the
approximation is good for all regimes in which the algorithms are
effective, the approximation breaking down at higher temperatures.

\section{Speed Tests}
\label{tests}

In the low $T$ limit, the average exit time from an $s=2$ MCAMC
algorithm iteration for the East model is approximately
$e^{2\beta}/(N_4d)^2$.  The corresponding average time step for
standard CT is $e^{\beta}/N_4 d$.  The $s=2$ time step becomes
larger by a factor of $e^{\beta}/N_4 d$.  It gets a speed-up from
$e^{\beta}$, and a slowdown from $N_4$, as the more excitations
that are present upon entering the algorithm the quicker it exits,
and from $d$, as the higher the dimension the more facilitated
sites are available.  At low temperature, however, $N_4 \approx N
\epsilon$, so for fixed system size $N$ the speed-up factor of
$s=2$ MCAMC with respect to CT grows as $e^{2 \beta}$.

In figures 1 and 2 we show speed tests comparing the performance of
the MCAMC algorithms to standard MC and CT on East model simulations.
Fig.\ 1 shows the temperature dependence of the CPU time required for
generating an equilibrium trajectory of total Monte Carlo time $10^7
\times e^{2\beta}$ in an East model of $N=10^5$ sites.  In an $s=1$ CT
algorithm the average CPU time for such a simulation is independent of
$T$.  Fig.\ 1 shows that at very high temperatures standard MC is the
fastest method, but as $T$ is lowered CT soon outperforms it.  At
lower temperatures $s=2$ MCAMC becomes more efficient than CT by a
approximately a factor of $e^{2\beta}$.  As the temperature is dropped
further, $s=7$ MCAMC provides a further improvement of approximately
$e^{2\beta}$, and so on.

\begin{figure}
  \includegraphics[width=\columnwidth]{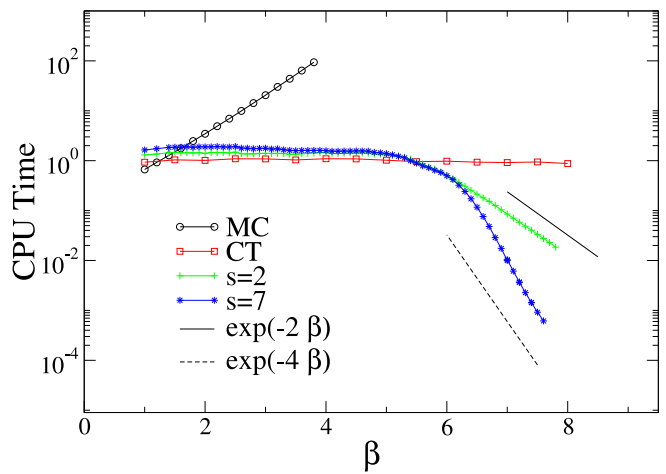}
  \caption{Temperature dependence of CPU time for equilibrium East
  model trajectories of total Monte Carlo time $t=10^7 \times e^{2\beta}$ and
  system size $N=10^5$, for MC, CT, and MCAMC algorithms.  The
  straight lines indicate the approximate speed-up of the MCAMC
  simulations.  CPU time shown relative to the average time needed
  when using CT dynamics.}
\end{figure}

\begin{figure}
  \includegraphics[width=\columnwidth]{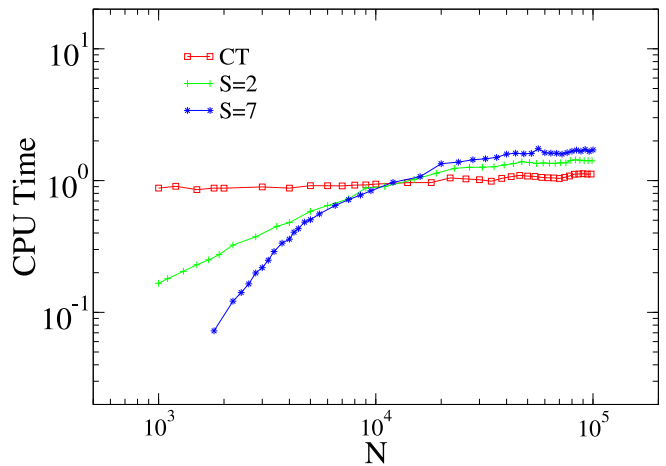}
  \caption{System size dependence of CPU time (relative to that for
  CT) for equilibrium East model trajectories of MC time
  $t=3 \times 10^{12}/N$ at $T=0.2$.}
\end{figure}

As discussed above, the efficiency of MCAMC depends on the system
size.  In addition to a reduced time step this also determines the
probability of encountering the isolated entry state for the $s>1$
algorithms. Fig.\ 2 shows the CPU time, now for different system
sizes, at fixed temperature and total MC time $t=3 \times
10^{12}/N$. Again, the CPU time for such a simulation using CT is
constant. As expected, Fig.\ 2 shows that as $N$ becomes larger
the MCAMC algorithms are less and less effective; beyond $N
\epsilon^2 \approx O(1)$ the CT scheme works better. This means
that in order to maximise the MCAMC efficiency one needs to
simulate the smallest possible system sizes.  This is limited by
the need to be compatible with bulk behaviour, which in the case
of facilitated models requires that the system in average contains
a sufficient number of excitations, i.e., $N \epsilon$ cannot be
too small.

\section{Example of results}
\label{results}

\begin{figure}
\includegraphics[width=\columnwidth]{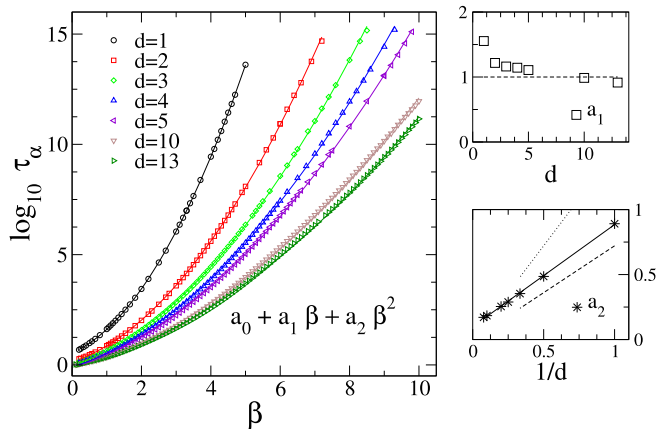}
\caption{Persistence time $\tau_\alpha$ as a function of inverse
  temperature $\beta=1/T$ in the East model in dimensions
  $d=1-5,10,13$.  The lines through the data points are quadratic
  fits, $\log \tau_\alpha = a_0 + a_1 \beta + a_2 \beta^2$, with $a_i$
  fitting parameters.  The fit suggests that $a_1 \sim 1$ in general.
  The bottom-right panel shows $a_2$ as a function of $d$.  This
  coefficient seems to go as $a_2 \approx b/d$, with the constant $b
  \approx 0.8$ (shown as a full line).  This value is between $\ln 2$
  (dotted line) and $(\ln 2)/2$ (dashed line). }
\end{figure}

In this section we present an example of numerical results obtained
with the MCAMC.

A useful correlation function to study the relaxation of facilitated
models is the persistence function $P(t)$,
e.g. \cite{BuhotG01,GarrahanC02,BerthierG05}, which gives the
probability that a site has not changed its state up to time $t$.  In
terms of the local persistence field $p_i(t)=0,1$, where $1$ indicates
that site $i$ has not flipped up to that time, and $0$ that it has
flipped at least once, the persistence function reads $P(t) = N^{-1}
\sum_i p_i(t)$.  In contrast to standard MC or CT simulations, the
MCAMC algorithm could run into problems when trying to measure
persistence.  By construction, it misses some of the events that could
occur whilst in the transient subspace, for example, from the isolated
state many spins could flip up and then flip back down before finally
exiting to an absorbing state.  At low temperatures in equilibrium,
however, the contribution of these events is negligible, as the vast
majority of changes to the global persistence function is from
existing excitations spreading out into unmoved territory, and this is
captured by the MCAMC algorithm.  In fact, the only sites one needs to
be concerned with are those immediately next to the initial
excitations, which are very few in equilibrium at low $T$.

Figure 3 shows the equilibrium persistence time
\cite{GarrahanC02,BerthierG03b}, $\tau_\alpha$, of the East model in
various dimensions $d$, calculated using the MCAMC algorithm with
$s=2$.  For all dimensions studied we find that $\tau_\alpha$ is a
super-Arrhenius function of $T$.  This seems to indicate that the East
model is a fragile in all dimensions.  Given that any simple
mean-field estimate of the relaxation in this model would give
Arrhenius behaviour, the above result would suggest that the East
model has no upper critical dimension to its dynamics
\cite{GarrahanC03}.  The data is compatible with $\log \tau_\alpha =
a_0 + a_1 \beta + a_2 \beta^2$, as expected if relaxation processes in
the East model in any $d$ are quasi one-dimensional
\cite{GarrahanC03}.  The coefficient $a_2$ of the quadratic fits is
compatible with $a_2 \approx b/d$ \cite{GarrahanC03}, with the
constant $b$ obeying $\ln 2 \ge b \ge (\ln 2)/2$, reminiscent of the
rigorous $d=1$ result of Ref.\ \cite{AldousD02}.  Note that the
timescales reached with the MCAMC in Fig.\ 3 are between three and
five orders of magnitude longer than in previous studies
\cite{BerthierG05}.

\begin{figure}
  \includegraphics[width=\columnwidth]{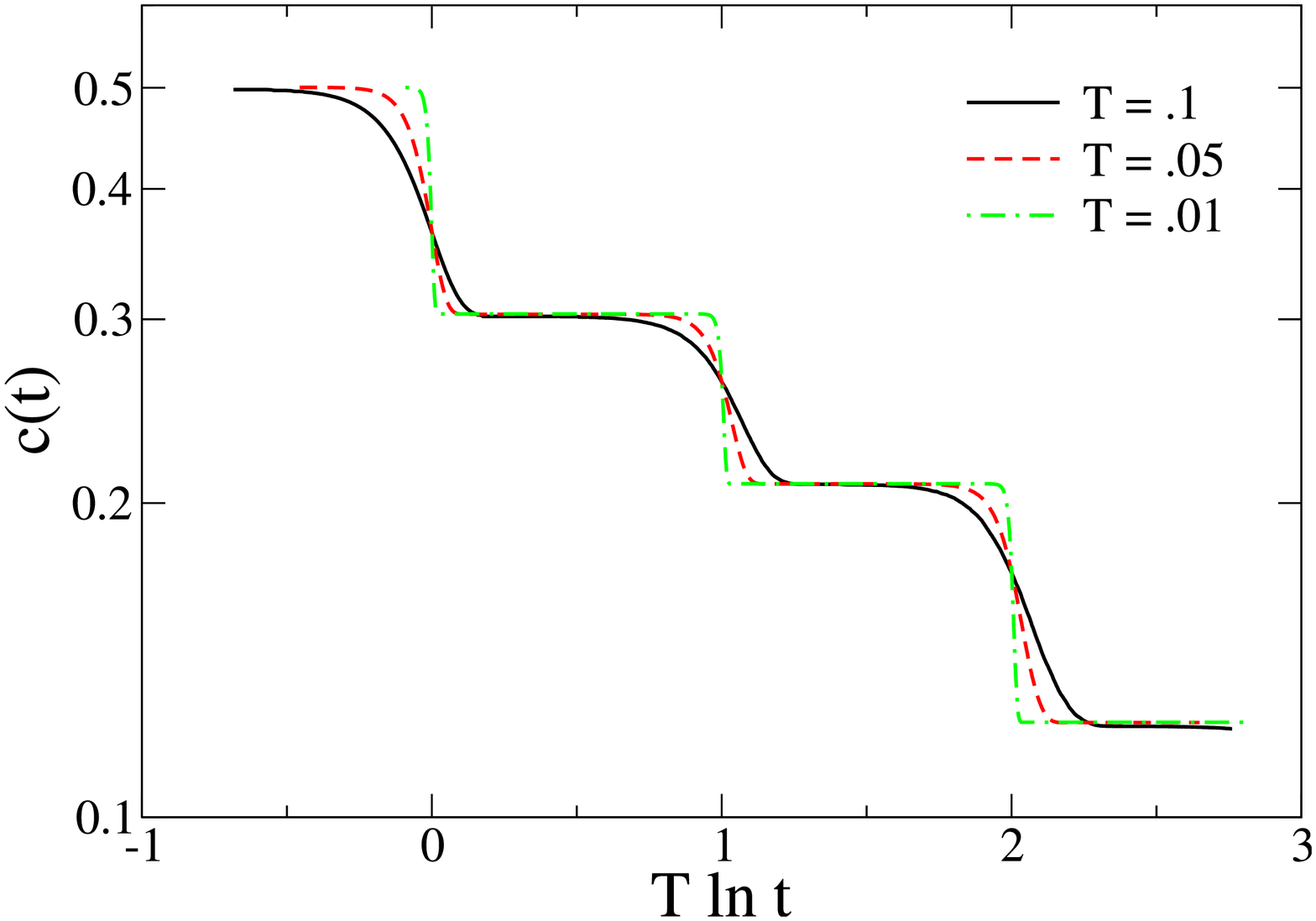}
  \caption{Concentration of excitations $c(t)$ as a function of scaled
  time $T \ln t$, in the $d=1$ East model after a quench from infinite
  temperature, from simulations with $s=7$ MCAMC. }
\end{figure}

The MCAMC proves also useful when simulating out-of-equilibrium
dynamics.  Consider the aging of the East model following a quench
from infinite temperature. As the system relaxes towards its
equilibrium the dynamics proceeds by stages characterized by the
distance between isolated excitations \cite{SollichE99}.  These
domains grow as $\overline{d} \sim t^{T \ln{2}}$.  Consequently, the
isolated transient state also plays an important role in such
out-of-equilibrium dynamics of the East model, and the MCAMC algorithm
is also applicable in this regime.  Figure 4 shows the aging of the
concentration of excitations, $c(t)$, with time after a quench to low
temperatures in the East model, using the $s=7$ MCAMC algorithm.

In these aging simulations the nature of the speed-up due to the MCAMC
becomes evident.  Each stage of the dynamics is associated with an
isolated domain of the form $10 \cdots 0$.  The $k$-th stage
corresponds to domains of typical length $l \sim 2^k$, and a
corresponding energy barrier of size $k$ to further relaxation
\cite{SollichE99}.  In essence, at each successive plateau of $c(t)$
one requires an algorithm that produces time steps comparable to the
activation time, $e^{k \beta}$.  An $s=2$ MCAMC enables one to push
simulations one plateau further than CT ($s=1$), the $s=7$ algorithm
helps overcome the next energy barrier, and so on.

\section{Discussion}
\label{conclu}

We have shown that the method of MC with absorbing Markov chains,
MCAMC, of Novotny \cite{Novotny95} can be used to dramatically speed
up simulations of facilitated spin models of glasses, such as the East
and FA models.  Even the simplest $s=2$ algorithm can improve
simulation times at low temperature by a factor of $e^{2 \beta}$ over
the n-fold or continuous time MC.  By increasing the number of
transient states $s$ even larger computational gains can be achieved.
One could imagine an algorithm where the number of transient states is
variable, and at each iteration the $s$ with the largest number of
transient states allowed by the current configuration, the smallest
distance between excitations, is used.

The next future step would be to adapt MCAMC algorithms to other
interesting KCMs, such as constrained lattice gases \cite{KobA93,
JackleK94,RitortS03,ToninelliBF04,PanGC04,LawlorDBSD05} and $f$-spin
facilitated FA models with $f>1$
\cite{FredricksonA84,RitortS03,SellittoBT05}.  Several features of
these systems make the application of MCAMC less straightforward:
since their kinetic constraints depend on more than one site,
i.e. facilitation by two or more excitations in the FA models or two
or more vacancies in the lattice gases, for generic entry states the
tree of possible transient states is much larger than for, say, East
models.  This means that the computational cost of the necessary
bookkeeping will be much higher (bookkeeping could be simplified by
reducing the possible entry states, at the expense of triggering less
frequently the MCAMC).  This problem is compounded by the fact that
$f>2$ FA models are very slow even at moderate temperatures, so that
the potential exponential in $\beta$ gains from excitation rates are
very modest, and may not even be enough to offset the bookkeeping
cost---in constrained lattice gases, where barriers are purely
entropic, this is bound to be worse.  In any case, given that the high
density or low temperature dynamics of these systems is in general so
much slower than that of East models, a clever MCAMC algorithm which
overcomes these hurdles could prove extremely useful.

\acknowledgements

We thank Robert Jack for discussions and Mark Novotny for
correspondence.  This work was supported by EPSRC grants no.\
GR/R83712/01 and GR/S54074/01, and University of Nottingham grant no.\
FEF 3024.


\end{document}